\documentclass[aps,prd
,nofootinbib,showpacs
,floatfix,superscriptaddress,preprintnumbers]{revtex4}
\usepackage{graphicx}
\usepackage{axodraw}
\usepackage{epsf}
\usepackage{epstopdf}
\usepackage{amsmath}
\usepackage{amssymb}
\usepackage{epsfig}
\usepackage{pstricks}
\usepackage{pst-coil}
\usepackage{verbatim}
\DeclareMathAlphabet{\mathsc}{OT1}{cmr}{m}{sc}

\newcommand{\beq}{\begin{equation}}
\newcommand{\eeq}{\end{equation}}
\newcommand{\bea}{\begin{eqnarray}}
\newcommand{\eea}{\end{eqnarray}}
\newcommand{\bec}{\begin{center}}
\newcommand{\eec}{\end{center}}
\newcommand{\bei}{\begin{itemize}}
\newcommand{\eei}{\end{itemize}}

\newcommand{\nn}  {\nonumber}

\def\10{$SO(10)$}
\def\21{SU(2) $\otimes$ U(1) }

\def\422{$SU(4) \otimes SU(2) \otimes SU(2)$}
\def\321{SU(3) $\otimes$ SU(2) $\otimes$ U(1)}

\def\lsim{\raise0.3ex\hbox{$\;<$\kern-0.75em\raise-1.1ex\hbox{$\sim\;$}}}
\def\gsim{\raise0.3ex\hbox{$\;>$\kern-0.75em\raise-1.1ex\hbox{$\sim\;$}}}
\def\vev#1{\left\langle #1\right\rangle}
\def\eq#1{eq.~(\ref{#1})}

\newcommand{\AddrAHEP}{%
  AHEP Group, Institut de F\'{\i}sica Corpuscular --
  C.S.I.C./Universitat de Val{\`e}ncia \\
  Edificio Institutos de Paterna, Apt 22085, E--46071 Valencia, Spain}

\newcommand{\ba}{\begin{array}}
\newcommand{\ea}{\end{array}}
\relax

\def\321{$SU(3)\times SU(2)\times U(1)$}

\begin{document}

\title{Quark mixing in the discrete dark matter model}

\author{Reinier de Adelhart Toorop$^{a}$, Federica Bazzocchi$^{b}$,
 Stefano Morisi$^{c}$ \\
\vspace{2mm}
$^{a}$~\it{Nikhef Theory Group,
Science Park 105, 1098 XG, Amsterdam, The Netherlands}\\
$^{b}$~\it{SISSA, Via Bonomea 265, Trieste, Italy and INFN, sez. Trieste}\\
$^{c}$~\it{\AddrAHEP}}


\date{\today}

\begin{abstract}
We consider a model in which dark matter is stable as it is charged under a $Z_2$ symmetry that is residual after an $A_4$ flavour symmetry is broken. We consider the possibility to generate the quark masses by charging the quarks appropriately under $A_4$. We find that it is possible to generate the CKM mixing matrix by an interplay of renormalisable and dimension-six operators. In this set-up, we predict the third neutrino mixing angle to be large and the dark matter relic density to be in the correct range. Low energy observables -- in particular meson-antimeson oscillations -- are hard to facilitate. We find that only in a situation where there is a strong cancellation between the Standard Model contribution and the contribution of the new Higgs fields, B meson oscillations are under control.
\end{abstract}

\maketitle

\section{Introduction}
Nowadays, there  is strong observational evidence of the existence of dark matter (DM)  \cite{Bertone:2004pz,Bertone}.
Many experiments are currently looking for direct or indirect observation of a dark matter candidate \cite{DMexp,Angle:2007uj,Ahmed:2009zw}. Among all the possible DM candidates stable cold dark matter (CDM) ones have been discussed in many   Standard Model (SM) extensions.  Typically its  stability  can be secured by introducing a parity, under which the CDM candidate is odd. This parity is often introduced \textit{ad hoc} with the sole purpose of making the DM candidate stable or it is related to a parity added to the theory by hand.

Recently, a model \cite{Hirsch:2010ru, Meloni:2010sk, Boucenna:2011tj} (to be referred to as discrete dark matter or DDM) was proposed that relates this DM parity to the residual symmetry of a spontaneously broken  flavour symmetry.  Flavour symmetries became popular after the discovery of particular patterns in neutrino mixing, since they can reproduce the observed structures from symmetry principles.

In DDM, only the lepton sector was studied. In this work, we will consider a simple way to add quarks to the model. We will show that this results in a diagonal CKM quark mixing matrix at the renormalisable level, but that non-renormalisable operators can generate correction to this in order to reproduce the observed mixing patterns.  It is not the first time that  a discrete lepton non-Abelian flavour symmetry is extended to the quark sector \cite{Altarelli:2010gt}-\cite{Ahn:2011yj}.  However, contrary to the majority of the cases in our scenario we do not have  \emph{flavons}, that is heavy flavour scalar SM singlets; instead SM  scalar doublets transform non trivially under the flavour symmetry. A consequence is the appearance of multiple new Higgs fields. Recently a similar set-up has been proposed in \cite{Meloni:2011cc}.

We will see that in our scenario  we can predict the third \textit{neutrino} mixing angle to be in the near-future experimental sensitivity (and indeed in the T2K range \cite{Abe:2011sj}) and that we have  a strong enhancement of meson-antimeson oscillations.  These are so strong that they rule out almost all parameter space for the model. Only if we allow a strong cancellation between the well-known contribution of the Standard Model and new contributions $\Delta M_{B_s}$, $\Delta M_{B_d}$ and $\Delta M_{K}$ can be reconciled with the data. Nevertheless we see only a  limited effect on the calculation of the relic density of the DM candidate.

\section{The model}
\label{model}

As in DDM, we assign matter fields to irreducible representations of $A_4$, the
group of even permutations of four objects, isomorphic to the symmetry
group of the tetrahedron. The group $A_4$ is often used in flavour model building as it naturally allows the neutrino mixing to be of the tribimaximal type \cite{Harrison:2002er}, that fits the observed data very well. The properties of the group $A_4$ are summarized in the appendix.

The representations of the leptons and the Higgs fields are as in DDM. In particular, there is an $SU(2)$-doublet Higgs $\hat{H}$ in the trivial representation of $A_4$ and we assume three extra copies of the Higgs $\eta=(\eta_1,\eta_2,\eta_3)$ transforming as a triplet. There are also four neutrinos: three of them transform as an $A_4$-triplet $N_T=(N_1,N_2,N_3)$ and a fourth transforms as a singlet $N_4$.
We assume that all quarks transform in the same way as the charged leptons. Both lefthanded and righthanded fields transform as one-dimensional representations of $A_4$, with different representations over the generations: the first generation is taken to transform as $1$, the second generation as $1'$ and the third generation as $1''$. They thus transform non-trivial under the $Z_3$ subgroup of $A_4$, but are uncharged under the $Z_2$ subgroup. All matter and Higgs assignments of our model are summarized in table\,\ref{tab1}.
\begin{table}[h!]
\begin{center}
\begin{tabular}{|c|c|c|c|c|c|c|c|c|c|c|c|c|c|c||c|c|}
\hline
&$Q_1$&$Q_2$&$Q_3$&$q_{R_1}$&$q_{R_2}$&$q_{R_3}$&$\,L_e\,$&$\,L_{\mu}\,$&$\,L_{\tau}\,$&$\,\,l_{Re}\,\,$&
$\,\,l_{{R\mu}}\,\,$&$\,\,l_{{R\tau}}\,\,$&$N_{T}\,$&$\,N_4\,$&$\,\hat{H}\,$&$\,\eta\,$\\
\hline
$SU(2)$&2&2&2&1&1&1&2&2&2&1&1&1&1&1&2&2\\
\hline
$A_4$ &1&$1'$&$1''$&1&$1'$&$1''$&$1$ &$1^\prime$&$1^{\prime \prime}$&$1$&$1^{\prime}$&$1''$&$3$ &$1$ &$1$&$3$\\
\hline
\end{tabular}\caption{Summary of  relevant  model quantum numbers. $q=u,d$.}\label{tab1}
\end{center}
\end{table}

The resulting Yukawa Lagrangian for the leptons is unchanged
\begin{equation}
\begin{split}
\label{laglep}
\mathcal{L}_{\rm l} = & \, y_e \overline{L}_el_{Re} \hat{H}+y_\mu \overline{L}_\mu l_{R\mu}^c \hat{H}+y_\tau \overline{L}_\tau l_{R\tau}^c \hat{H} +\\
& y_1^\nu \overline{L}_e(N_T\tilde{\eta})_{1}+y_2^\nu  \overline{L}_\mu(N_T\tilde{\eta})_{1''}+y_3^\nu  \overline{L}_\tau(N_T\tilde{\eta})_{1'}+\\
& y_4^\nu  \overline{L}_e N_4 \hat{H}+ M_1 N_TN_T+M_2 N_4N_4 +
\mbox{h.c.},
\end{split}
\end{equation}
where we have defined $\tilde{\eta}$ as $i \sigma_2 \eta^*$.

For the quarks, the Lagrangian reads
\begin{equation}
\begin{split}
\label{lagqua}
\mathcal{L}_{\rm q} = & \, y_u \overline{Q}_1 \tilde{H} u_{1R}+y_c \overline{Q}_2 \tilde{H} u_{2R}+y_t \overline{Q}_3 \tilde{H} u_{3R} + \\
& y_d \overline{Q}_1 \hat{H} d_{1R}+y_s \overline{Q}_2 \hat{H} d_{2R}+y_b \overline{Q}_3 \hat{H} d_{3R}+
\mbox{h.c.},
\end{split}
\end{equation}
where $\tilde{H}$ stands for $\tilde{\hat{H}}$, defined analogously to $\tilde{\eta}$.

At the renormalisable level, both up- and downquark matrices are diagonal with all masses given by $m_i = y_i v_H/\sqrt{2}$, with $v_H/\sqrt{2}$ the vacuum expectation value of the $A_4$-singlet Higgs field. As we do not aim to explain the hierarchy of the quark masses, with $m_u \approx 10^{-5} m_t$, we assume an hierarchy in the Yukawa couplings, with most Yukawa coupling being small to very small. For instance a Froggatt-Nielsen symmetry \cite{FN} could make this more natural.

As discussed in more detail in \cite{Hirsch:2010ru},  electroweak symmetry is broken by the vacuum configuration
\begin{equation}
\label{vev}
\vev{\hat{H}^0}= v_H / \sqrt{2},\qquad \vev{ \eta^0_1}=(v_\eta / \sqrt{2},0,0) \,.
\end{equation}
We write the ratio between the vev of $H$ and $\eta$ as $\tan \tilde{\beta}$ and obviously, their squares sum to (246 GeV${})^2$. The vev of $\eta$ breaks the  $A_4$ group into its subgroup $Z_2$, generated by $S$, that is diagonal in the three-dimensional representation: $S=\text{Diag}(1,-1,-1 )$. The $Z_2$
symmetry thus acts on the $A_4$  triplet fields in the following way:
\begin{equation}\label{residualZ2}
Z_2:\quad
\begin{array}{lcrlcrlcr}
N_2 &\to& -N_2\,,\quad& h_2 &\to& -h_2\,, \quad&A_2 &\to& -A_2\,, \\
N_3 &\to& -N_3\,,\quad& h_3 &\to& -h_3\,,\quad &A_3&\to& -A_3\,,
\end{array}
\end{equation}
where $N_{2,3}$ are the components of the triplet $N_T$ and $h_{2,3}$ and $A_{2,3}$ are respectively the CP-odd and CP-even components of the Higgs doublet $\eta_{2,3}$.

The residual $Z_2$ symmetry is responsible for the stability of the lightest combination of $h_2$, $
h_3$, $A_2$ and $A_3$ which is the dark matter candidate.
Indeed the $Z_2$-odd candidate only couples to heavy right-handed neutrinos and not to the SM charged fermions, that are $Z_2$-even. Such a  scalar dark matter candidate is potentially detectable in nuclear  recoil experiments \cite{Angle:2007uj,Ahmed:2009zw}.

We refer to the four $Z_2$-even components of the $Z_2$-even Higgs fields $\hat{H}$ and $\eta_1$ as
$H'_0$, $H'_1$, $A'_0$ and $A_1'$ in accordance with \cite{Boucenna:2011tj}. They give rise to two scalars $H$ and $H_0$, one pseudoscalar $A_0$ and the neutral Goldstone boson of electroweak symmetry breaking.

As mentioned above, the charged leptons and quarks transform non-trivially under the $Z_3$ subgroup of $A_4$ generated by $T$. The vev configuration (\ref{vev}) clearly breaks this $Z_3$. Still, at the tree level, quark and charged lepton masses preserve the $Z_3$ invariance thanks to the  scalar charge assignments. As we will see in the next section $Z_3$ breaking effects appear at next to leading order  (NLO) level giving rise to the quark mixing matrix.

\section{Quark mixing}
Quark masses at the tree level are given by equation (\ref{lagqua}). This gives rise to diagonal quark mass matrices and the CKM matrix $V_{\textrm{CKM}} = (V_L^u)^\dagger V_L^d$ is simply the identity matrix.

The gauge and flavour charge assignments in the Higgs sector allow the construction dimension six operators for the down-type quark masses that contain $\hat{H}$, $\eta$ and their conjugates. There are three ways to contract the $SU(2)$ indices, represented by brackets in the equation below
\begin{equation}\label{o6}
\sum \frac{f_{ij} }{\Lambda^2}(\overline{Q}_{i} \hat{H}) d_{j} (\eta^\dagger \eta)+ \frac{f'_{ij} }{\Lambda^2} (\overline{Q}_{i}  \eta) d_{j}  (\eta^\dagger \hat{H}) +  \frac{f''_{ij} }{\Lambda^2} (\overline{Q}_{i}  \eta) d_{j}  (\hat{H}^\dagger \eta)\,.
\end{equation}
The contraction of $A_4$ indices between the two $\eta^{(\dagger)}$ triplets (see equation (\ref{3x3product})) is such that it generates the right type of singlet (1, 1' or 1'') to match the charges for $\overline{Q}_{i}$ and $d_{j}$. It is important to note that this is possible for any combination of $i$ and $j$ due to the product rules of $A_4$. $\Lambda$ is the cut-off scale, up to which we accept the theory to be valid and the $f$ couplings are dimensionless. Analogous dimension-6 operators can obviously be constructed for up-type quarks and charged leptons.

The mass term Lagrangian (\ref{lagqua}) and the effective couplings (\ref{o6}) generate the effective mass matrix for down-type quarks
\begin{equation}
\label{Md}
M_d=\left(
\begin{array}{ccc}
m_d & 0 & 0\\
0 & m_s & 0 \\
0&0& m_b
\end{array}
\right)+
\frac{v_H v_\eta^2}{\Lambda^2}
\left(
\begin{array}{ccc}
h_{dd}&h_{ds}&h_{db}\\
h_{sd}&h_{ss}&h_{sb}\\
h_{bd}&h_{bs}&h_{bb}
\end{array}
\right)+\mathcal{O}(1/\Lambda^4)\,,
\end{equation}
where $h_{ij}= (f_{ij}+f'_{ij}+f''_{ij})/2\sqrt{2}$.
Analogous expressions again hold  for up-type quarks and charged leptons.

Now the crucial question is how large the cut-off scale $\Lambda$ is. In principle, this is a scale we are free to set. Only using `naturalness' and `finetuning' arguments, we can find a range for it. We will give two arguments, both pointing to a scale of 1 to 10 TeV.

In the first argument, we demand that there should not be more than 10 to 100\% corrections to the Higgs from one-loop corrections to the Higgs propagator with the fermions and the (new) scalars of the theory. These corrections are typically of the order $\Lambda^2 / (4 \pi)^2$ and requiring them to be not too large with respect to $v_{ew}^2 = (246 \textrm{GeV})^2$ indeed gives $\Lambda \lesssim (1 \textrm{ to } 10) \textrm{ TeV}.$

Interestingly, we find the same scale from an argument where we require the dimensionless parameters $h$, in particular $h_{ds}$, to be of order 1. The off-diagonal terms in equation (\ref{Md}) are responsible for generating the quark mixing as parameterized by the CKM matrix. As we do not have information about the size of the dimensionless parameters $h$, we assume them to be of order 1, which can be seen as the most natural assumption for dimensionless parameters.

Under this assumption, the absolute values of the corrections to the leading order elements of the mass matrix are of the same order for the up-type quark matrix and the down-type quark matrix. However, due to the much larger elements of the leading order up-type quark mass matrix, the effects on quark mixing are dominated by the down-type quark contributions. This allows us to estimate the order of magnitude of the cut-off scale.

Now the (1 2) element of equation (\ref{Md}) should be of order $\lambda_C \, m_s$ in order to reproduce the Cabibbo angle.
\beq
h_{ds}   \frac{v_H v_\eta^2}{\Lambda^2} = \lambda_C m_s.
\eeq
This gives
\beq
\Lambda^2 = h_{ds} \frac{v_H v_\eta^2}{\lambda_C m_s} =  h_{bd} \frac{v_{ew}^3}{(\tan^2 \tilde{\beta}) (1+\frac{1}{\tan^2 \tilde{\beta}})^{3/2} \lambda_C m_s} = [ (1 \textrm{ to } 10) \textrm{ TeV} ]^2,
\eeq
depending on the exact values of $h_{ds}$ and $\tan \tilde{\beta}$, which we have taken between 0.1 and 1 and between 0.1 and 10 respectively. Due to the large bottom mass, the effect of  $h_{db}$ and $h_{sb}$ of the same size as $h_{ds}$  on $\theta_{13}$ and $\theta_{23}$ is relatively minor and these angles are thus naturally smaller than the Cabibbo angle and indeed in a large part of parameter space, we can fit them to their measured values.

The analogue of the dimension 6 operator (\ref{o6}) affects the lepton mixing. In \cite{Hirsch:2010ru}, it was shown that at leading order, the lepton mixing  matrix has zero $\theta_{13}$-angle and large $\theta_{12}$ and $\theta_{23}$ mixing angles, although these do not necessarily fit in a mixing pattern such as tribimaximal or bimaximal mixing. The fact that the down-type quark and charged lepton mass matrices are alike (at least at leading order) suggests that the matrices that diagonalize them, $V_L^d$ and $V_L^e$ are also similar. We thus expect a large angle (of the order of the Cabibbo angle) in the (1 2) sector of $V_L^e$ and in the lepton mixing matrix $(V_L^e)^\dagger \, V_L^\nu$, this affects all three angles. In particular, we expect a Cabibbo-sized correction to the $\theta_{13}$-angle. This is in the range of values hinted at by the recent T2K results \cite{Abe:2011sj} and in recent global fits \cite{Fogli:2011qn}. In any case, we predict that non-zero $\theta_{13}$ should clearly be measured by next generation experiments, such as Daya Bay \cite{dayabay} and Double Chooz \cite{doublechooz}.

We end this section with a comment on the scale of the neutrino seesaw. In the DDM model, neutrino masses are assumed to originate from the type-I seesaw
\beq
m_\nu= -m_{D_{3\times 4}}M_{R_{4\times 4}}^{-1}m_{D_{3\times 4}}^T =
-\frac{v_\eta^2}{2 M_1}
\left(
\begin{array}{ccc}
(y_1^\nu)^2 + (y_4^\nu)^2 \frac{M_1}{M_2}& y_1^\nu \, y_2^\nu  & y_1^\nu \, y_3^\nu\\
y_1^\nu \, y_2^\nu & (y_2^\nu)^2 & y_2^\nu \, y_3^\nu\\
y_1^\nu \, y_3^\nu & y_2^\nu \, y_3^\nu & (y_3^\nu)^2
\end{array}
\right).
\eeq
This has two non-zero eigenvalues, that are of the order $y_i^\nu \, y_j^\nu \, v_\eta^2 / M_k$. There can be a low energy seesaw, where we identify the mass scale of the righthanded neutrinos to $\Lambda$ if the neutrino Yukawa couplings are not too small $y_\nu = 10^{-(4\div5)}$. Lastly, next-to-leading order effects lift the mass of the lightest neutrino away from zero. Still, it is suppressed with respect to the other neutrino masses by a factor $v_\eta / \Lambda$.

\section{FCNC}

The inclusion of the operators \eq{o6} gives rise to tree level FCNC processes mediated by the $Z_2$-even scalar  ($H_0$ and $H$) and pseudoscalar  ($A_0$). In the mass  eigenstate basis, the trilinear couplings read
\bea
1/(2 \sqrt{2})\bar{F}_{Li} f_{Rj} H v_\eta^2 v_h ([ U_{{HH_{0}}}]_{ 11}/v_h  + 2 [U_{{HH_{0}}}]_{ 21}/v_\eta ) (f_{ij}+f'_{ij}+f''_{ij})\,,&&\nn\\
&& \nn\\
1/(2 \sqrt{2})\bar{F}_{Li} f_{Rj} H_0 v_\eta^2 v_h ([U_{{HH_{0}}}]_{ 12}/v_h  + 2 [U_{{HH_{0}}}]_{ 22}/v_\eta ) (f_{ij}+f'_{ij}+f''_{ij})\,,&&\nn\\
&& \nn\\
  \pm 1/(2 \sqrt{2}) \bar{F}_{Li} f_{Rj}  A_0[ U_{G A_{0}}]_{ 12} (f_{ij}+f'_{ij}-f''_{ij})/v_h+  2    [U_{G A_{0}}]_{ 22}   f''_{ij}/v_\eta\,,&&
\eea
 where in the last equation $+$ is for up-quarks and $-$ for down-quarks--and charged leptons as well.
$U_{HH_0}$ and $U_{G A_{0}}$ are the matrices that relate the  scalar and pseudoscalar mass and interaction eigenstates; in particular   $U_{G A_{0}}$  is a rotation matrix over the angle $\tilde{\beta}$ \cite{Boucenna:2011tj}.

\begin{figure}[ht!]
\begin{center}
\includegraphics[height=4cm]{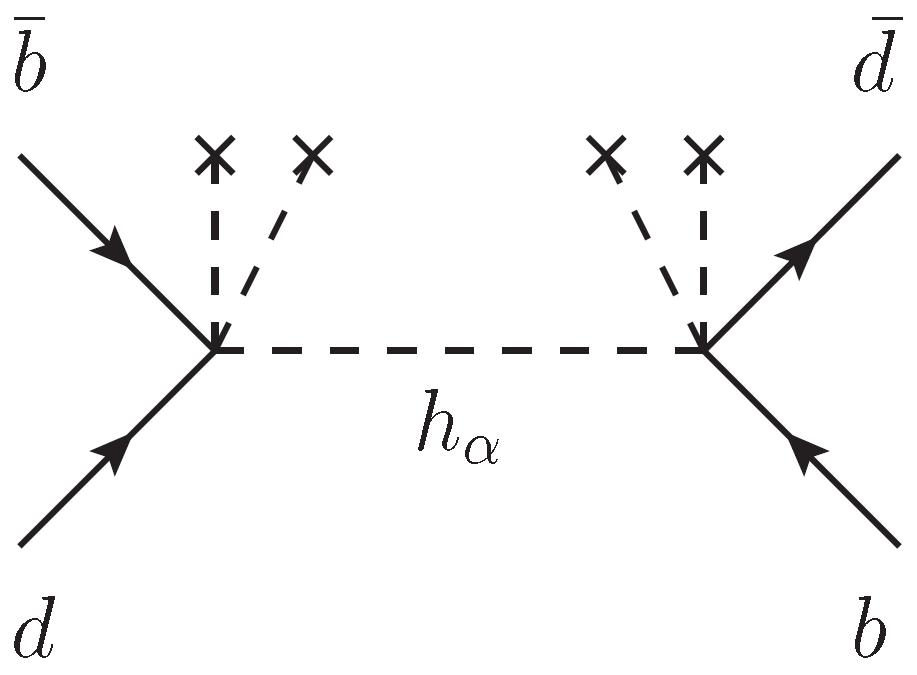} \\
\caption{\label{fig2}\it Feynman diagrams for B mesons oscillations in the SM (left) and in our model (right); similar diagrams can be drawn for B${}_s$,  and K mesons.}
\end{center}
\end{figure}

In this section, we will focus on meson-antimeson oscillations, as these are among the most constraining tests for new physics. In particular, we find that they are more constraining than the often-discussed K-meson decays. In our model, the Standard Model box diagrams are accompanied by new tree diagrams; see figure \ref{fig2}.
In models with multiple Higgs bosons, the new physics (NP) contribution to the mass splitting connected to $F^0 - \overline{F}^0$ oscillations is given by \cite{Toorop:2010kt, Atwood:1996vj, Wells:2009kq}
\beq
\Delta M_F^{\mathrm{NP}} = B_F^2 \ f_F^2 \ M_F \sum_\alpha \bigg[ \frac{1}{m_H^\alpha{}^2} \Big( |I^\alpha_{rs}|^2 \Big(\frac{1}{6} + \frac{1}{6} \frac{M_F^2}{(m_r + m_s)^2} \Big) +  |J^\alpha_{rs}|^2 \Big(\frac{1}{6} + \frac{11}{6} \frac{M_F^2}{(m_r + m_s)^2} \Big) \Big) \bigg].
\eeq
Here, $M_F$ is the mass of the meson, $f_F$ is its decay constant and $B_F$ are recalibration constants of order 1, related to vacuum insertion formalism. The masses, $m_r$ and $m_s$ are those of the quarks of which the meson is build, i.e. $rs = bd, \, bs, \, ds$ stands for $B_d, \, B_s$ and $K^0$ respectively. Lastly, $I^\alpha_{rs}$ and $J^\alpha_{rs}$ are effective fermion-fermion-scalar and fermion-fermion-pseudoscalar couplings, as given in \cite{Toorop:2010kt}.

\begin{figure}[ht!]
\begin{center}
\includegraphics[width=8cm]{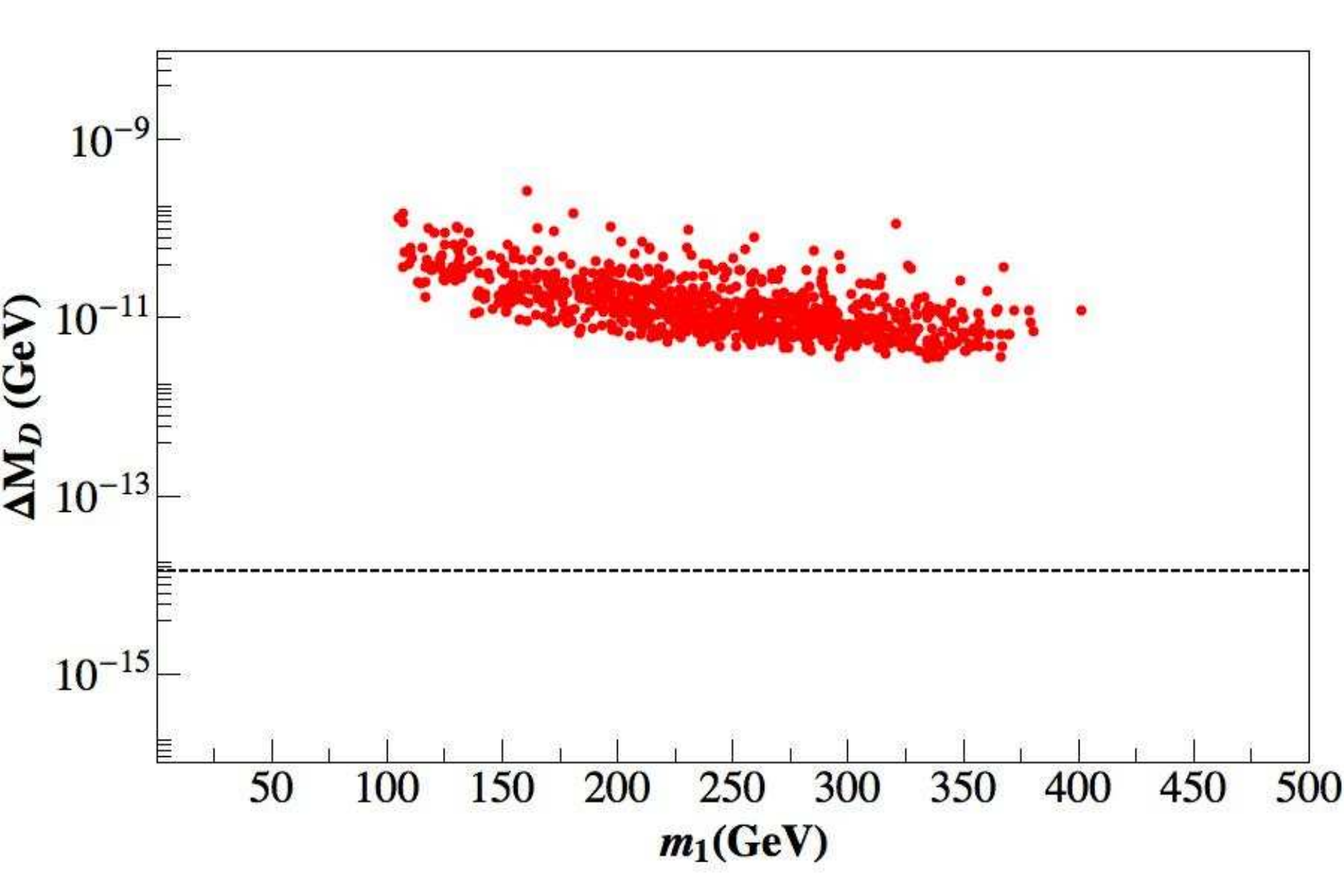} \\
\caption{\label{figD}\it D meson oscillations in a scenario where the CKM matrix is generated dominantly by corrections to the up-type quark mass matrix.}
\end{center}
\end{figure}

Indeed we find that the meson-antimeson oscillations severely reduce the parameter space of our model. As mentioned in the previous section, the CKM matrix can dominantly originate from corrections to the up-type quark mass matrices or the down-type quark mass matrices and we mentioned that dominance of the latter is more natural. Indeed we find that if corrections of the former type dominate or even if there is no dominance of one of the two, $\Delta M_D$ of D meson oscillations is much larger than the experimental value \cite{Wells:2009kq} and this scenario should be excluded as  shown in figure \ref{figD}.

\begin{figure}[ht!]
\begin{center}
\includegraphics[width=8cm]{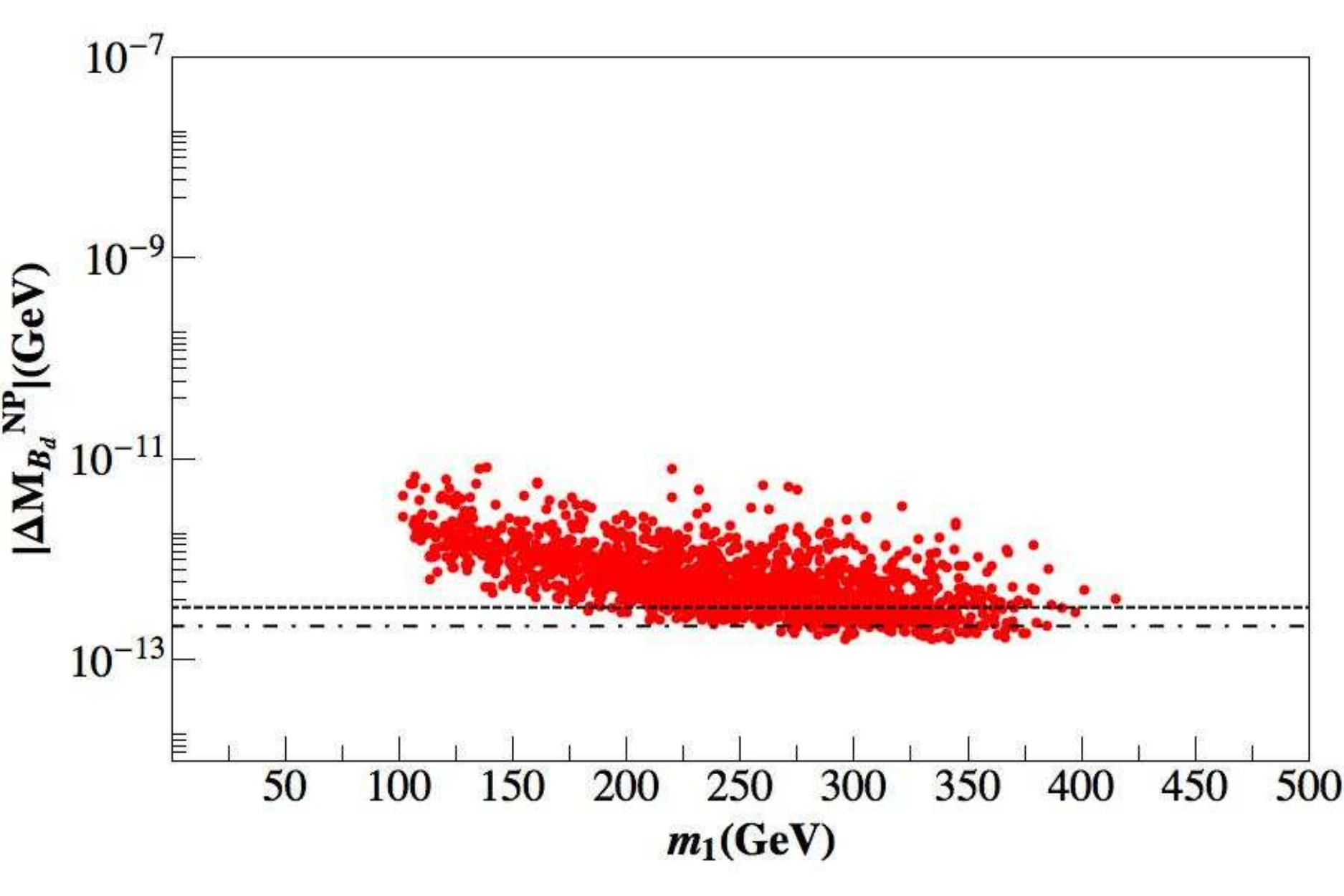}
\includegraphics[width=8cm]{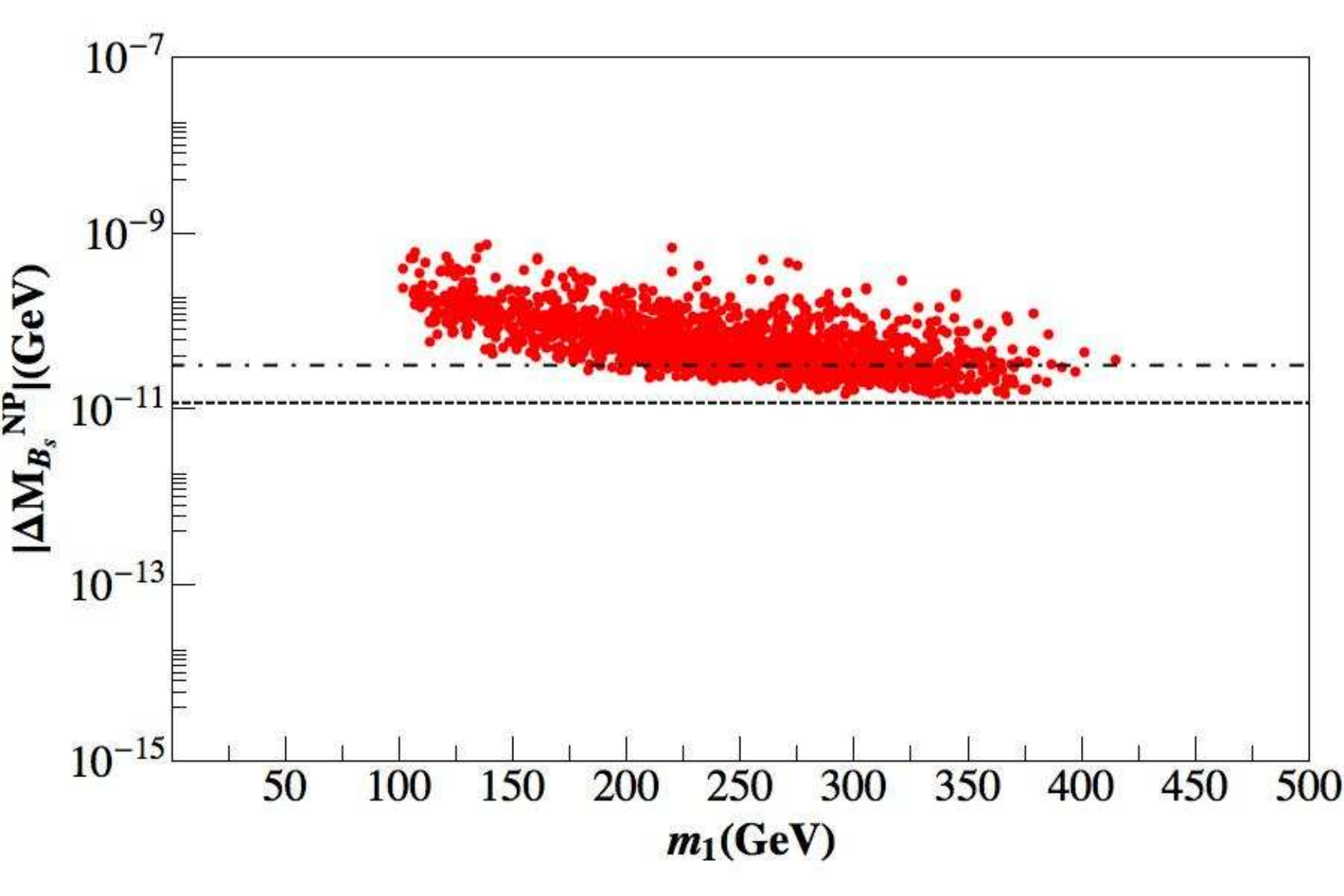}\\
\includegraphics[width=8cm]{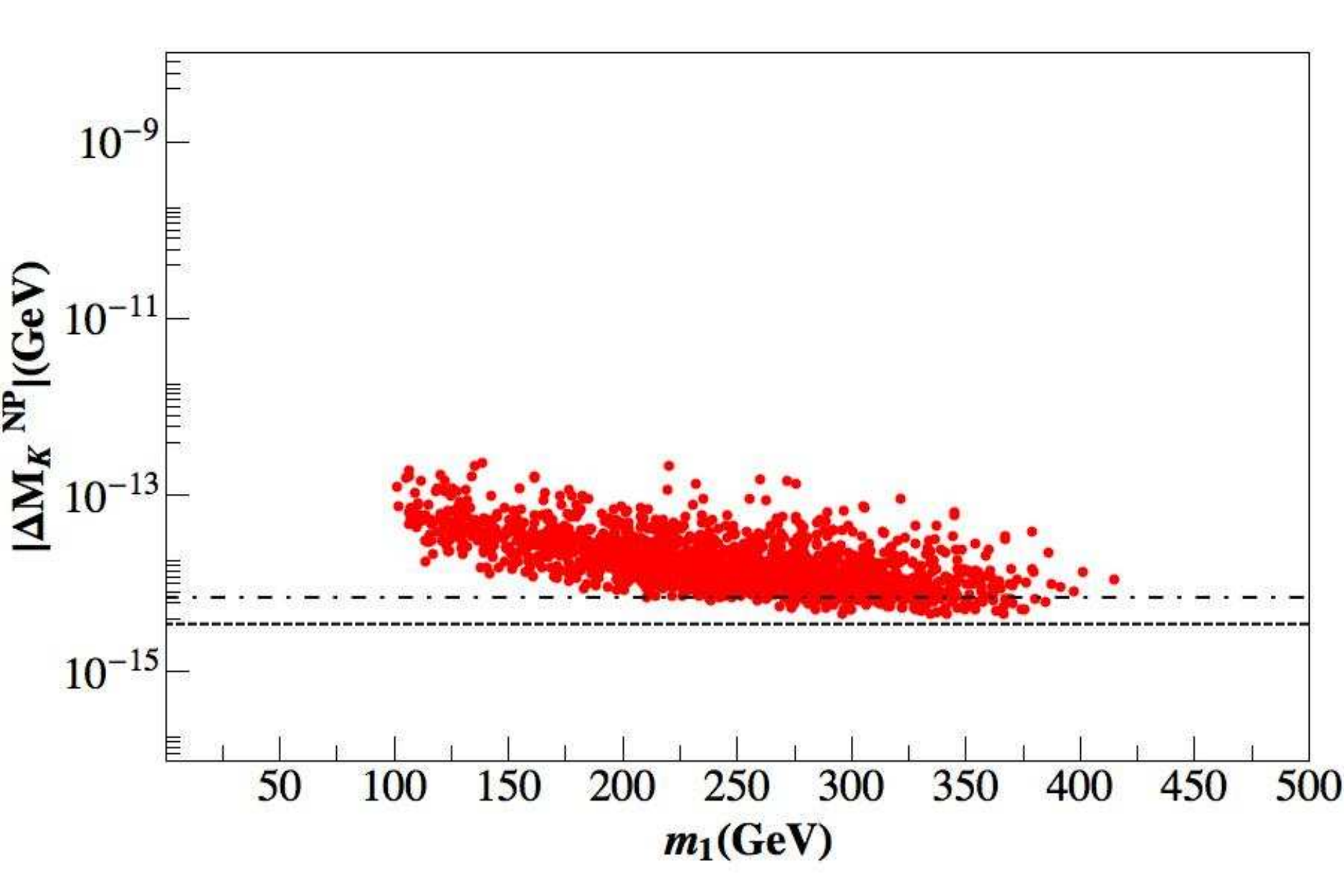}
\caption{\label{figMdplot}\it B and K mesons oscillations. The horizontal lines are as explained in the text.}
\end{center}
\end{figure}

In case the CKM matrix mostly originates from corrections to the down-type quark mass matrix, we find that the bounds from meson mixing are rather strong. In figure \ref{figMdplot}, we show the contribution of the new diagrams to these as well as to $\Delta M_{B_d}$ as function of the lightest Higgs mass. We see that in almost all of parameter space the points are near or even above the short-dashed line, which indicates the current experimental value \cite{Buras:2010mh} that is rather well described by the Standard Model box diagrams \cite{Nakamura:2010zzi, Dubinin:2007qt}. Naively, this is interpreted as an exclusion of the model, which is true in most of parameter space, but not in points where the Standard Model and new physics contributions partially cancel. To see this, we write the mixing amplitude for $B_d$ mixing as \cite{Botella:2006va}
\beq
\big[ M_{12}^d \big]_{\mathrm{NP}} = (1 - \frac{1}{1+h_d^2 e^{2i \sigma_d}}) \big[ M_{12}^d \big]_{\mathrm{full}}
\eeq
and analogously for $B_s$. We have verified that our expression for the NP contribution carries enough phases to generate a flat distribution for $\sigma_d$ and $\sigma_s$. We check that the points in $(h_d^2 = 0.41, \sigma_d = 100^\circ)$ and $(h_s^2 = 1.6, \sigma_s = 90^\circ)$ are allowed by the data \cite{Botella:2006va, Grossman:2009dw} and give a nett contribution of NP with respect to the full result of respectively 0.65 and 2.67 times the observed value. For $B_d$ mesons, NP effects are forced to be less than the observed value, while for $B_s$ mesons, it can be slightly more. These values (and a corresponding estimate in case of the kaons) correspond to the dot-dashed lines in \ref{figMdplot}. We see that a small, but significant number of points is allowed by the data under the assumption of partial cancellation.

The naturalness of this cancellation, that requires finetuning between the phase and amplitude of the new and the Standard Model contribution, can be a matter of debate. The need for cancellation diminishes for larger Higgs masses, although there are still no points below the lower line for $B_s$ and $K$. Indeed, requiring that the new diagrams contribute less than the experimental bound, as is customarily done, the model would be excluded.

\begin{figure}[ht!]
\begin{center}
\includegraphics[width=16cm]{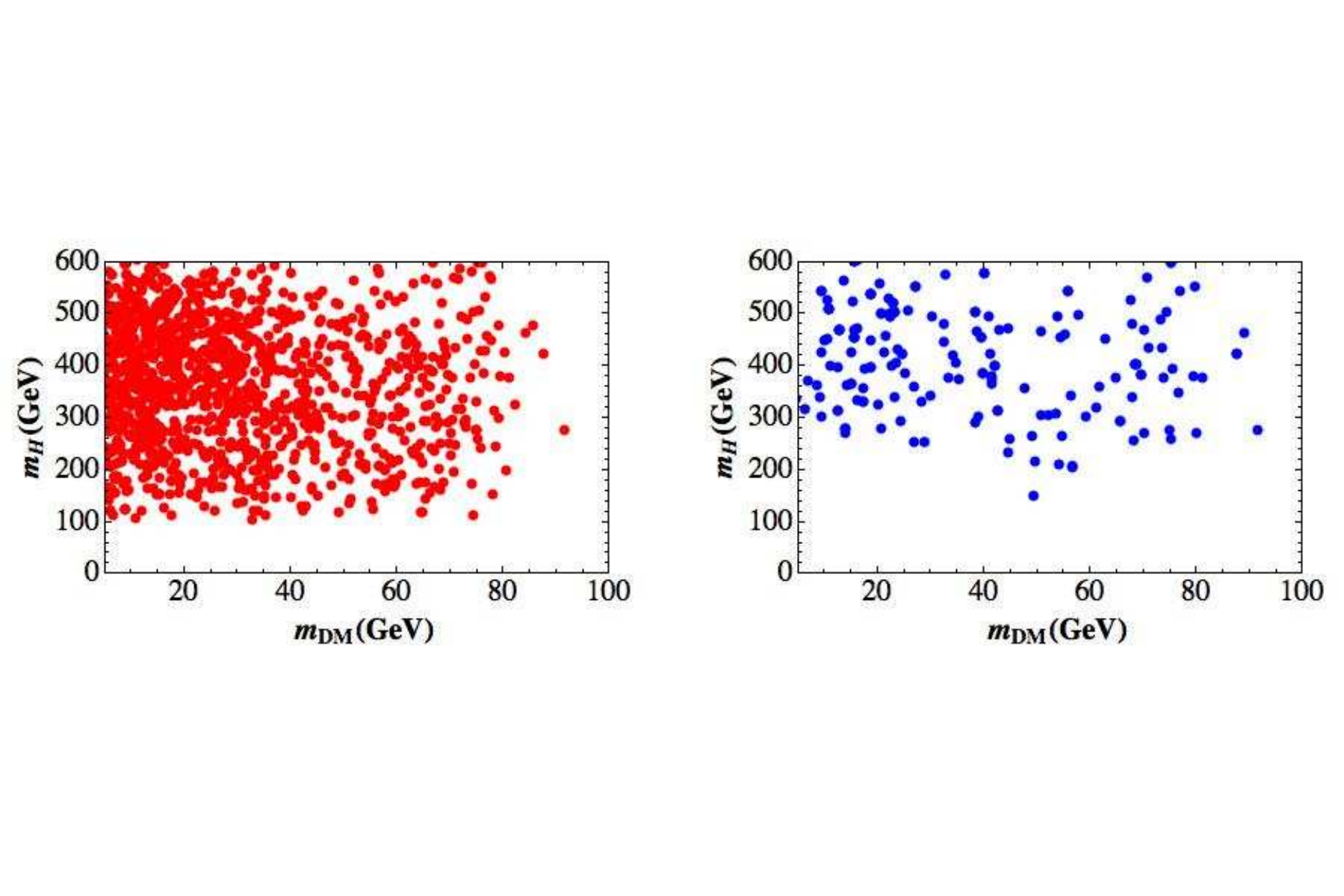}
\caption{\label{Higgsmasses}\it  The mass of the lightest $Z_2$-even state versus the mass of the DM candidate without (red) and with (blue) meson oscillation constraints.}
\end{center}
\end{figure}

On the other hand allowing a strong negative interference between the SM  and the DDM contributions does not further constrain the scalar spectrum with respect to the analysis done in \cite{Boucenna:2011tj}. Indeed fig. \ref{Higgsmasses} shows that there is no correlation between the  bound imposed and the mass of the lightest  $Z_2$-even scalar state, even if the number of points allowed significantly reduces with respect to those in  \cite{Boucenna:2011tj}.

\section{Dark matter relic density and direct detection}

The dark matter candidate of the DDM model can annihilate into up or down quarks via SM Higgs exchange in the s-channel. The operators (\ref{o6}) give rise to a new contribution for this decay in the case of decay into down quarks. Decay to up quarks is not affected as we have seen that up-type quarks are forced to be almost diagonal. In  \cite{Boucenna:2011tj}
it was shown that the model --without NLO terms -- gives rise to an available DM candidate.
The relic dark matter density constrains the parameter space of the model. In particular it was found
that dark matter mass $M_{DM}$ is in the range 1 - 100 GeV. The region with $M_{DM}<40$\,GeV and mass
of the Standard Model Higgs $M_H>400$\,Gev is excluded, while for $M_{DM}>50$\,GeV the Higgs mass can go up to about
$500$\,GeV as co-annihilation can be possible. For $M_{DM}$ lighter than about $80$\,GeV the dark matter annihilate
(coannihilate) into fermions trough exchange of scalar (pseudoscalar and gauge boson) in the s-channel. For
masses heavier than about $80$\,GeV the main channels of annihilation are with $W$ bosons in the final state.

Below we consider  the effect of the inclusion of the dimension-six terms  of eq.\,(\ref{o6}) to the relic density.
The operators\,(\ref{o6}) give an effective quartic coupling of the dark matter with quarks. We therefore study
the effects of such a operator only for dark matter mass below $80$\,GeV as for heavier dark matter mass the main annihilation is into $W$ bosons and not in quarks.

We recall that the DM candidate is one of the four neutral states (2 scalar and 2 pseudoscalar) arising by
the mixing of the neutral components of $\eta_2$ and $\eta_3$ that in \cite{Boucenna:2011tj} were indicated
as $H'_m+i  A'_m$ with $m=2,3$.
From eq.\,(\ref{o6}) we derive the $H'_m,A'_m$ coupling to fermions. For the couple $H'_m H'_m$
with $m=2,3$ the coupling is proportional to
\begin{equation}
v_H(f_{ij}+f'_{ij}+f''_{ij})/\Lambda^2= v_H h_{ij}/\Lambda^2\,,
\end{equation}
while for  $A'_m A'_m$  it is proportional to
\begin{equation}
v_H(f_{ij}+f'_{ij}-f''_{ij})/\Lambda^2\,.
\end{equation}

If the DM candidate turns out to be the CP-even $Z_2$-odd  lightest state we may estimate the contribution
of the new operator to the total $\sigma_{\chi \chi\to d_i\bar{d_j}}$. Similar conclusions would be obtained considering the case in which the DM candidate is the CP-odd $Z_2$-odd state.
In the previous section we have seen that
\begin{equation}
h _{ij}^{d} v_{H}\frac{v_\eta^2}{\Lambda^2}\sim m_s\lambda_C\,,
\end{equation}
for any couple $(ij) \neq 11$,  since for the first family the $h^{d}_{11}$ is required to be smaller to fit down mass. There we need
\begin{equation}
h _{11}^{d} v_{H}\frac{v_\eta^2}{\Lambda^2}\sim m_d\,.
\end{equation}
We define the $\lambda^q_{ eff}$  as the parameter of the four points interaction $\chi \chi d_i\bar{d}_j$. We can estimate its size as
\begin{equation}
\lambda^d_{eff}\sim
h _{ij}^{d} \frac{v_H}{\Lambda^2}\sim \frac{m_s \lambda_C}{v_\eta^2}\,.
\end{equation}
We compare the effect of this operator on the the $\sigma v_{rel}$ of the process $\chi \chi\to d_i\bar{d}_i$ to the effect of $s$-channel exchange of the SM-like lightest CP-even scalar.
The SM-like Higgs couples to fermions proportionally to $y_q \sim m_q/v_{H}$ and it turns out that  the new contribution is negligible if
\begin{equation}
\label{DM1}
\frac{m_s \lambda_C}{v_\eta^2}< \frac{m_q A_H}{v_H m_h^2}\,.
\end{equation}
In \eq{DM1} $m_h$  is the mass of the lightest CP-even neutral scalar and $A_H\sim v_W$  is the dimensional coupling that controls the
interaction of the dark matter with the Higgs doublet $ H \chi \chi$.
Since $ v_\eta\sim v_H\sim m_h\sim A_H \sim v_W$ the new contribution   is naturally subleading for the second and third generation.
For the first  generation, on the contrary the new contribution  to $\chi \chi\to d\bar{d}$ is of the same order as the old one. We can (conservatively) estimate that this channel is negligible only for for values $M_{DM}\geq1$\,GeV.

NLO terms also give rise to non-diagonal scattering $\chi \chi\to d_i\bar{d}_j$, which was not included in the  analysis done in \cite{Boucenna:2011tj}. In particular the scattering   $\chi \chi\to d \bar{b}$ can be non-negligible for DM masses around a GeV. For this reason we conclude that  the previous analysis is not affected  in the range
$M_{DM}\geq5$\,GeV. We take this lower bound as a further constraint on the scalar sector parameter space, postponing a complete  new analysis  to the future \cite{noi}.

Direct detection is not affected at all by the NLO terms: the quark flavour  diagonal scattering contribution are subdominant with respect to  those mediated by the scalar $H$ while the off-diagonal one could only give rise to processes that are not kinematically allowed such as  $ \chi + \mathcal{N} \to \chi +\mathcal{N} + \pi^+ + e^- +\bar{\nu}$, with $\mathcal{N}$ a nucleus in the detector bulk.

\vskip5.mm

\section{Conclusions}
Neutrino mixing might be explained by a discrete non-Abelian  flavour symmetry. If this symmetry is dynamically broken only in one direction,   a residual symmetry survives. It is interesting that this residual symmetry may be responsible for the existence of  a stable dark matter candidate. It has already been shown that this set up can describe the physics in the lepton sector and the dark matter abundance rather well.

It is a natural question to see if the quark sector can also be described in such scenarios. In this paper we investigated this possibility in a particular model \cite{Hirsch:2010ru, Meloni:2010sk, Boucenna:2011tj}. We found that if we add quarks in the same non-trivial representations of the flavour symmetry as charged leptons, the interplay between renormalizable operators and dimension-six operators can generate a realistic CKM matrix in a large portion of parameter space. It is possible  to let the CKM matrix dominantly originate from the up-type quark sector as well as the down-type quark sector, although the latter case is more natural.

The new dimension six operators should also be present in the lepton sector. As a consequence we have shown that the predictions of the original model, $\theta_{13}=0$ and $m_3=0$, are shifted away from zero, in the former case into the near-future observable region.  At the same time for what concerns the quark sector  the new operators can lead to new channels for flavour changing neutral currents and we have analysed their effects on meson-antimeson oscillations. We found that D meson oscillations rule out the scenario where the CKM matrix originates mostly from the up sector and that K and B${}_{s,d}$ meson oscillations are much enhanced as well, with some of their amplitudes at least as large as corresponding the Standard Model amplitudes, bringing the model close to being excluded, although a certain number of points in parameter space can still be reconciled with current observations. If this is the case  the scalar sector behaves exactly as  described in \cite{Boucenna:2011tj}.

Although these observations do not rule out the extension of the discrete dark matter model with quarks, severe finetuning is needed and honesty forces us to say that this  `natural extension' is less natural than we hoped.

\section*{Acknowledgments}
We thank M. Taoso  for his help in the numerical analysis and P. Ullio for helpful discussion. SM's work was supported by the Spanish MICINN under grants
FPA2008-00319/FPA and MULTIDARK Consolider CSD2009-00064, by
Prometeo/2009/091, by the EU grant UNILHC PITN-GA-2009-237920
and by a Juan de la Cierva contract. The work of RdAT  is part of the research program of the Dutch Foundation for Fundamental Research of Matter (FOM). RdAT acknowledges the hospitality of the University of Padova, where part of this research was completed.

\appendix

\section{The group $A_4$}

All 24 elements of $A_4$ are generated from two elements
$S$ and $T$ with $S^2=T^3=(ST)^3=\mathcal{I}$.
$A_4$ has four irreducible representations, three singlets
$1,~1^\prime$ and $1^{\prime \prime}$ and one triplet $3$.

We can choose a basis in which $S$ and $T$ can be represented as $(1, 1)$, $(1,\omega)$, $(1,\omega^2)$, with $\omega = e^{2 \pi i /3}$ for the three one-dimensional representations and
\begin{equation}\label{eq:ST}
S=\left(
\begin{array}{ccc}
1&0&0\\
0&-1&0\\
0&0&-1\\
\end{array}
\right)\,;\quad
T=\left(
\begin{array}{ccc}
0&1&0\\
0&0&1\\
1&0&0\\
\end{array}
\right)\,;
\end{equation}
for the three-dimensional representation. $S$ and $T$ on themselves generate the two maximal subgroups of $A_4$, the Abelian $Z_2$ and $Z_3$. The $A_4$ multiplication rules are given by
\begin{equation}
\begin{split}
&1 \times r = r \textrm{ for all representations $r$,} \\
&1' \times 1' = 1'', \quad 1'' \times 1'' = 1, \quad 1' \times 1'' = 1, \\
&1' \times 3 = 3, \quad 1'' \times 3 = 3,\\
&3 \times 3 =  1 + 1'+ 1'' + 3 + 3.
\end{split}
\end{equation}
Representing the two triplets as $a=(a_1,a_2,a_3)$ and $b=(b_1,b_2,b_3)$, the elements of the last product are
\begin{equation}
\label{3x3product}
\begin{array}{lll}
(ab)_1&=&a_1b_1+a_2b_2+a_3b_3\,;\\
(ab)_{1'}&=&a_1b_1+\omega a_2b_2+\omega^2a_3b_3\,;\\
(ab)_{1''}&=&a_1b_1+\omega^2 a_2b_2+\omega a_3b_3\,;\\
(ab)_{3_1}&=&(a_2b_3,a_3b_1,a_1b_2)\,;\\
(ab)_{3_2}&=&(a_3b_2,a_1b_3,a_2b_1)\,.
\end{array}
\end{equation}


\begin{thebibliography}{9}

\bibitem{Bertone:2004pz}
  G.~Bertone, D.~Hooper and J.~Silk,
  Phys.\ Rept.\  {\bf 405}, 279 (2005)
  [arXiv:hep-ph/0404175].

\bibitem{Bertone}
Bertone, G. (ed.) ÒParticle dark matter: Evidence, candidates and constraintsÓ, (Cambridge Univ. Press, 2010).

\bibitem{DMexp}
Papers of WMAP observations available at: /http://lambda.gafc.nasa.gov/ product/map/current/map-bibliography.cfmS; R. Bernabei, 804 (2008) astro-ph/ 08042741; Pamela Collaboration, O. Adriani, et al., Nature 458 (2009).

\bibitem{Angle:2007uj}
  J.~Angle {\it et al.} [ XENON Collaboration ],
  Phys.\ Rev.\ Lett.\  {\bf 100 } (2008)  021303.
  [arXiv:0706.0039 [astro-ph]].

\bibitem{Ahmed:2009zw}
  Z.~Ahmed {\it et al.} [ The CDMS-II Collaboration ],
  Science {\bf 327 } (2010)  1619-1621.
  [arXiv:0912.3592 [astro-ph.CO]].

\bibitem{Hirsch:2010ru}
  M.~Hirsch, S.~Morisi, E.~Peinado and J.~W.~F.~Valle,
  Phys.\ Rev.\  D {\bf 82} (2010) 116003
  [arXiv:1007.0871 [hep-ph]].


\bibitem{Meloni:2010sk}
  D.~Meloni, S.~Morisi and E.~Peinado,
  Phys.\ Lett.\  B {\bf 697} (2011) 339
  [arXiv:1011.1371 [hep-ph]].


\bibitem{Boucenna:2011tj}
  M.~S.~Boucenna, M.~Hirsch, S.~Morisi, E.~Peinado, M.~Taoso and J.~W.~F.~Valle,
  arXiv:1101.2874 [hep-ph].

\bibitem{Altarelli:2010gt}
  G.~Altarelli and F.~Feruglio,
  Rev.\ Mod.\ Phys.\  {\bf 82}, 2701 (2010)
  [arXiv:1002.0211 [hep-ph]].


\bibitem{Feruglio:2007uu}
  F.~Feruglio, C.~Hagedorn, Y.~Lin and L.~Merlo,
  Nucl.\ Phys.\  B {\bf 775}, 120 (2007)
  [Erratum-ibid.\  {\bf 836}, 127 (2010)]
  [arXiv:hep-ph/0702194].


\bibitem{Everett:2010rd}
  L.~L.~Everett and A.~J.~Stuart,
  Phys.\ Lett.\  B {\bf 698}, 131 (2011)
  [arXiv:1011.4928 [hep-ph]].

\bibitem{Frampton:2007et}
  P.~H.~Frampton and T.~W.~Kephart,
  JHEP {\bf 0709}, 110 (2007)
  [arXiv:0706.1186 [hep-ph]].


\bibitem{Chen:2007afa}
  M.~C.~Chen and K.~T.~Mahanthappa,
  Phys.\ Lett.\  B {\bf 652}, 34 (2007)
  [arXiv:0705.0714 [hep-ph]].





\bibitem{Babu:2004tn}
  K.~S.~Babu and J.~Kubo,
  Phys.\ Rev.\  D {\bf 71}, 056006 (2005)
  [arXiv:hep-ph/0411226].

\bibitem{Blum:2007nt}
  A.~Blum, C.~Hagedorn and A.~Hohenegger,
  JHEP {\bf 0803}, 070 (2008)
  [arXiv:0710.5061 [hep-ph]].

\bibitem{Blum:2007jz}
  A.~Blum, C.~Hagedorn and M.~Lindner,
  Phys.\ Rev.\  D {\bf 77}, 076004 (2008)
  [arXiv:0709.3450 [hep-ph]].




\bibitem{Dong:2010zu}
  P.~V.~Dong, H.~N.~Long, D.~V.~Soa and V.~V.~Vien,
  Eur.\ Phys.\ J.\  C {\bf 71}, 1544 (2011)
  [arXiv:1009.2328 [hep-ph]].


\bibitem{Toorop:2010yh}
  R.~de Adelhart Toorop, F.~Bazzocchi and L.~Merlo,
  JHEP {\bf 1008}, 001 (2010)
  [arXiv:1003.4502 [hep-ph]].


\bibitem{Hagedorn:2010th}
  C.~Hagedorn, S.~F.~King and C.~Luhn,
  JHEP {\bf 1006}, 048 (2010)
  [arXiv:1003.4249 [hep-ph]].


\bibitem{Ding:2009iy}
  G.~J.~Ding,
  Nucl.\ Phys.\  B {\bf 827}, 82 (2010)
  [arXiv:0909.2210 [hep-ph]].


\bibitem{Bazzocchi:2009pv}
  F.~Bazzocchi, L.~Merlo and S.~Morisi,
  Nucl.\ Phys.\  B {\bf 816}, 204 (2009)
  [arXiv:0901.2086 [hep-ph]].


\bibitem{Ishimori:2008fi}
  H.~Ishimori, Y.~Shimizu and M.~Tanimoto,
  Prog.\ Theor.\ Phys.\  {\bf 121}, 769 (2009)
  [arXiv:0812.5031 [hep-ph]].


\bibitem{Ding:2008rj}
  G.~J.~Ding,
  Phys.\ Rev.\  D {\bf 78}, 036011 (2008)
  [arXiv:0803.2278 [hep-ph]].


\bibitem{Morisi:2011pt}
  S.~Morisi, E.~Peinado, Y.~Shimizu and J.~W.~F.~Valle,
  arXiv:1104.1633 [hep-ph].


\bibitem{Lavoura:2007dw}
  L.~Lavoura and H.~Kuhbock,
  Eur.\ Phys.\ J.\  C {\bf 55}, 303 (2008)
  [arXiv:0711.0670 [hep-ph]].


\bibitem{Bazzocchi:2007au}
  F.~Bazzocchi, S.~Morisi and M.~Picariello,
  Phys.\ Lett.\  B {\bf 659}, 628 (2008)
  [arXiv:0710.2928 [hep-ph]].

\bibitem{Bazzocchi:2007na}
  F.~Bazzocchi, S.~Kaneko and S.~Morisi,
  JHEP {\bf 0803}, 063 (2008)
  [arXiv:0707.3032 [hep-ph]].

\bibitem{Ahn:2011yj}
Y.~H.~Ahn, H.~Y.~Cheng and S.~Oh,
Phys.\ Rev.\  D {\bf 83}, 076012 (2011)
[arXiv:1102.0879 [hep-ph]].

\bibitem{Meloni:2011cc}
  D.~Meloni, S.~Morisi and E.~Peinado,
  arXiv:1104.0178 [hep-ph].

\bibitem{Abe:2011sj}
  K.~Abe {\it et al.} [ T2K Collaboration ],
  Phys.\ Rev.\ Lett.\  {\bf 107 } (2011)  041801.
  [arXiv:1106.2822 [hep-ex]].

\bibitem{Harrison:2002er}
  P.~F.~Harrison, D.~H.~Perkins and W.~G.~Scott,
  Phys.\ Lett.\  B {\bf 530} (2002) 167
  [arXiv:hep-ph/0202074].

\bibitem{FN}
C.~D.~Froggatt and H.~B.~Nielsen,
  {\it Hierarchy Of Quark Masses, Cabibbo Angles And CP Violation},
  Nucl. Phys. B {\bf 147} (1979) 277.

\bibitem{Fogli:2011qn}
  G.~L.~Fogli, E.~Lisi, A.~Marrone, A.~Palazzo, A.~M.~Rotunno,
  Phys.\ Rev.\  {\bf D84 } (2011)  053007.
  [arXiv:1106.6028 [hep-ph]].

\bibitem{dayabay}
Y.~f.~Wang,
  arXiv:hep-ex/0610024;
X.~Guo {\it et al.}  [Daya-Bay Collaboration],
  arXiv:hep-ex/0701029.

\bibitem{doublechooz}
F.~Ardellier {\it et al.},
  arXiv:hep-ex/0405032;
F.~Ardellier {\it et al.} [Double Chooz Collaboration],
  arXiv:hep-ex/0606025.


\bibitem{Atwood:1996vj}
  D.~Atwood, L.~Reina, A.~Soni,
  Phys.\ Rev.\  {\bf D55 } (1997)  3156-3176.
  [hep-ph/9609279].

\bibitem{Wells:2009kq}
  J.~D.~Wells,
  [arXiv:0909.4541 [hep-ph]].

\bibitem{Toorop:2010kt}
  R.~de Adelhart Toorop, F.~Bazzocchi, L.~Merlo, A.~Paris,
  JHEP {\bf 1103 } (2011)  040.
  [arXiv:1012.2091 [hep-ph]].

\bibitem{Buras:2010mh}
  A.~J.~Buras, M.~V.~Carlucci, S.~Gori, G.~Isidori,
  JHEP {\bf 1010 } (2010)  009.
  [arXiv:1005.5310 [hep-ph]].


\bibitem{Nakamura:2010zzi}
  K.~Nakamura {\it et al.} [ Particle Data Group Collaboration ],
  J.\ Phys.\ G {\bf G37 } (2010)  075021.

\bibitem{Dubinin:2007qt}
  M.~Dubinin, A.~Sukachev,
  Phys.\ Atom.\ Nucl.\  {\bf 71 } (2008)  374-387.
  [arXiv:0711.5023 [hep-ph]].

\bibitem{Botella:2006va}
  F.~J.~Botella, G.~C.~Branco, M.~Nebot,
  Nucl.\ Phys.\  {\bf B768 } (2007)  1-20.
  [hep-ph/0608100].

\bibitem{Grossman:2009dw}
  Y.~Grossman, Z.~Ligeti, Y.~Nir,
  Prog.\ Theor.\ Phys.\  {\bf 122 } (2009)  125-143.
  [arXiv:0904.4262 [hep-ph]].

\bibitem{noi}
Work in progress.
\end{thebibliography}
\end{document}